# Mini Review on the Significance Nano-Lubricants in Boundary Lubrication Regime

Mohamed Kamal Ahmed Ali [*,1,2], Mohamed A. A. Abdelkareem [1,2]

Ahmed Elagouz [1,2], F. A. Essa [3], Hou Xianjun [*,1]

[1] Hubei Key Laboratory of Advanced Technology for Automotive Components, Wuhan University of Technology, Wuhan 430070, China. [2] Automotive and Tractors Engineering Department, Faculty of Engineering, Minia University, El-Minia 61111, Egypt. [3] School of Mechanical and Electronic Engineering, Wuhan University of Technology, Wuhan 430070, China.

[*]Corresponding Author:

eng.m.kamal@mu.edu.eg (M.K.A. Ali) & houxj@whut.edu.cn (H. Xianjun)

**Abstract**

This article briefly reviews the significance nano-lubricants and assesses their effectiveness to provide the most promising approaches to reduce the friction and anti-wear/scuffing over the boundary regime. The main purpose of this review is to summarize the present knowledge about major advantages of the nanomaterials as nanolubricant additives in boundary lubrication. It is very complex regime involving surface topography, metallurgy, physical adsorption and chemical reactions. There is no intent to present an exhaustive survey of the literature, but it presents the main reasons for decreasing the friction and wear based on lubricated by nanolubricants during heavy load and low-speed conditions.

**Introduction**

Stribeck curve describes the friction levels of contacts with different film thickness to surface roughness ratios with lubrication regimes of the major components in automotive engines such as piston rings, engine bearings and valves as shown in Figure 1. Internal combustion Engine lubrication is categorized into three main regimes: boundary, mixed and elastohydrodynamic lubrication [1]. In piston rings/liner assembly, the previous lubrication regimes can be obtained over the stroke depending on operating conditions. Generally, boundary lubrication exists under the effect of low speed and high load conditions [2]. Hence, nano-lubricant additives are very important in boundary lubrication because of the higher friction coefficient [3]. Current challenges for reducing the friction and wear require an adaptable lubricant for different operating conditions. Thence, most researchers have focused on nano-lubricant concept in the internal combustion engines as the main strategy for suppressing the friction coefficient and the wear of contact surfaces, in a manner that will ultimately lead to an improved tribological performance [4, 5]. The main advantages of the nanoparticles additives compared to conventional lubricant additives are stated as follows [6]:

1- Nanoparticles are often efficient at room temperature.

2- Activation of nanoparticle surface.

3- Increase of surface area and extreme small sizes.

4- Nanoparticles sized smaller than 100 nm have thermal conductivity higher than of the fluids.

5- Nanoparticles provide excellent tribological performance as solid lubricant.

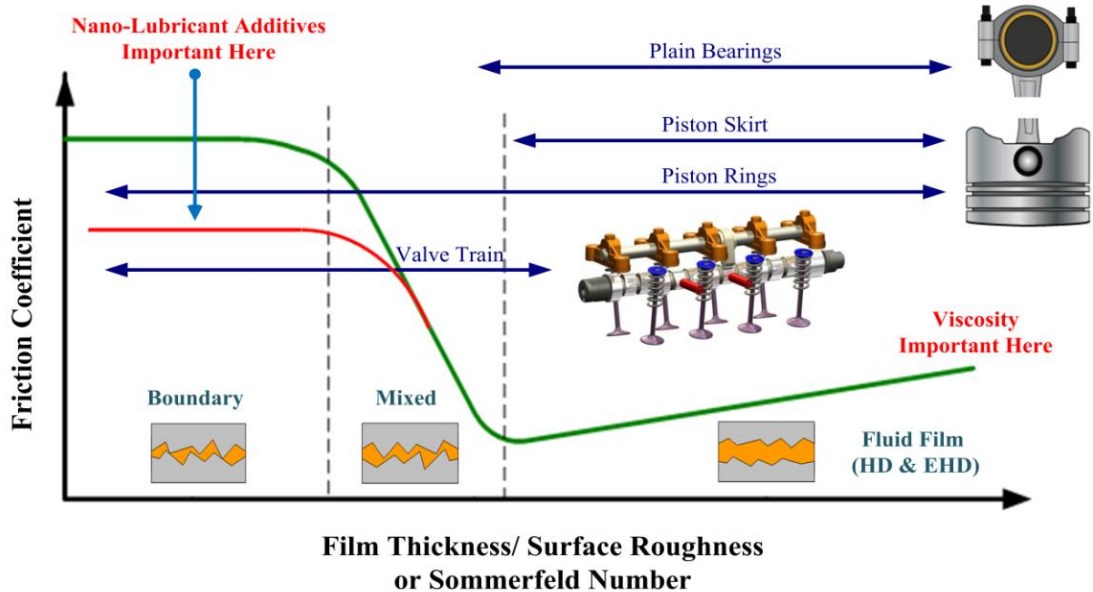

**Figure 1.** Lubrication stribeck curve of the major components in an internal combustion engines.

**Discussion**

The chemical transformation of both engine lube oil and lubrication additives during operation are known to be quite complicated. Currently, minimizing the total frictional power losses in vehicle engines is a persistent challenge and most of the recent efforts have focused on enhancing the anti-wear and anti-friction by nanomaterials as nano-lubricant additives. Hence, decreasing boundary friction coefficient between the rings and liner plays a critical role in improving engine performance and fuel economy. Notably, The load is supported by the surface peaks not by the lubricant film. The experimental results by Ali et al. [4] showed that the friction coefficient during boundary lubrication decreased by 35-51% near ends of the stroke (TDC and BDC) for the $Al_2O_3$ and $TiO_2$ nano-lubricants, respectively. The significant decline in the friction coefficient might be because of the leading role of nanomaterials to act through different mechanisms such as micro rolling bearing, tribofilm deposition, and mending effect mechanisms. Additionally, the spherical morphology of the nanoparticles used as nanoparticles lube additives is also considered an important factor that supplies a rolling effect between rubbing surfaces. Because of this reason, nanolubricants are most effective under boundary lubrication conditions (Fig. 1). When the grain size of the nano-additives is smaller than the irregular surface roughness, the valleys between the frictional surfaces asperities can be stuffed with the nano-additives leading to formulate a tribo-boundary film on the worn surface that can enhance the tribological behavior [1].

Ali et al. [7] described the reduced frictional power losses and anti-scuffing in automotive engines using $Al_2O_3/TiO_2$ nano-lubricant additives. The obtained results showed that the frictional power losses for the simulated ring/liner assembly, and wear rate of the piston ring dropped by 40-51% and 17% after sliding distance of 50 km, respectively. Moreover, the experimental fabricated tests showed that the friction coefficient decline with increasing the contact load and sliding speed for both nano-lubricant and lube oil without nano-additives. It is known from tribological experiments that the separation distance between the worn surfaces grows with increasing sliding speed. Thus, there is less time for the individual asperity contacts and, consequently, less time for asperities to deform. This usually results in a decrease of the real contact area and a decrease of the friction coefficient with increasing sliding speeds. While, the main causes for the decrease in wear rate are the high sliding speeds of the worn surfaces, which lead to thermal activation of the surfaces as well as the nanoparticles and thus helps to formulate a self-assembled tribofilm as a lubricating layer on the asperities causing less metal-to-metal contact as stated by [8].

Under high loads, the severity of contact pressure over these asperity tips could be enormous. At the same time, some of the asperities may fracture or crush, which results in the reduction of the boundary friction coefficient and smoother surface. Furthermore, frictional heating of asperity tips can also accelerate the tribofilm formation by using $Al_2O_3/TiO_2$ as a hybrid nanoparticle additive. This can be translated into more effective tribofilm deposition on the rubbing surfaces as a solid lubricant and thus suppresses the wear of the ring under the higher loading condition. Hence, it has been noted that the anti-scuffing effect increased owing to the formation of a self-tribofilm layer, which formed on the rubbing surfaces as a solid lubricant. Ali et al. [9] studied the effect of both $Al_2O_3$ and $TiO_2$ nanoparticles additive into lube oil on the thermophysical parameters. The thermophysical parameters are thermal conductivity and viscosity. The results presented that the $Al_2O_3/TiO_2$ nanolubricants provide low kinematic viscosity and an increase in the viscosity index by 2%. Moreover, thermal conductivity was enhanced by 12-16%, compared with the case of engine oil without nanoparticles.

**Conclusion**

The principal motivation for formulating new additives using nanoparticles as a promising solution for improving the tribological behavior is that nanoparticles have the potential to offer significant tribological benefits of both solid and liquid lubrication and extend the life of the mechanical components. The decline of the friction coefficient between the rings/liner assembly plays a critical role in improving engine performance and fuel efficiency. Nanolubricant additives play a significant role in the formation of a tribofilm layer on the worn surfaces via physical or chemical absorbed mechanism to enhance protection of the worn surfaces and create a rolling effect between sliding surfaces.

**Acknowledgments**

The authors would like to express their deep appreciation to the Hubei Key Laboratory of Advanced Technology for Automotive Components (Wuhan University of Technology) for finacial support.

**References**

[1] Ali MKA, Xianjun H, Mai L, Qingping C, Turkson RF, Bicheng C. Improving the tribological characteristics of piston ring assembly in automotive engines using $Al_2O_3$ and $TiO_2$ nanomaterials as nano-lubricant additives. Tribology International. 2016;103:540-54.

[2] Ali MKA, Xianjun H, Turkson RF, Ezzat M. An analytical study of tribological parameters between piston ring and cylinder liner in internal combustion engines. Proceedings of the Institution of Mechanical Engineers, Part K: Journal of Multi-body Dynamics, 2016; 230, 329-349.

[3] Ali MKA, Xianjun H. Improving the tribological behavior of internal combustion engines via the addition of nanoparticles to engine oils. Nanotechnology Reviews. 2015;4:347-58.

[4] Ali MKA, Xianjun H, Elagouz A, Essa F, Abdelkareem MA. Minimizing of the boundary friction coefficient in automotive engines using $Al_2O_3$ and $TiO_2$ nanoparticles. Journal of Nanoparticle Research. 2016;18:377.

[5] Ali MKA, Ezzat FMH, 侯献军, 陈必成, 蔡清平.

边界润滑条件下机油污染物对摩擦学性能的影响. *润滑与密封* 42.2 (2017): 1-5.

[6] Martin JM, Ohmae N. Nanolubricants, Colloidal Lubrication: General Principles, John Wiley & Sons; 2008, pp. 1-3.

[7] Ali MKA, Xianjun H, Mai L, Bicheng C, Turkson RF, Qingping C. Reducing frictional power losses and improving the scuffing resistance in automotive engines using hybrid nanomaterials as nano-lubricant additives. Wear. 2016;364:270-81.

[8] Stott FH. The role of oxidation in the wear of alloys. Tribology International. 1998;31:61-71.

[9] Ali MKA, Xianjun H, Turkson RF, Peng Z, Chen X. Enhancing the thermophysical properties and tribological behaviour of engine oils using nano-lubricant additives. RSC Advances. 2016;6:77913-24.